\documentclass[12pt]{article}
\usepackage{graphicx}

\newcommand{\dzero}{D^0}
\newcommand{\dzerobar}{\overline{D}{}^0}
\newcommand{\zdmix}{$D^0$-$\overline{D}{}^0$~}
\newcommand{\taudzero}{\tau_{D^0}}

\def\Title#1{\begin{center} {\Large #1 } \end{center}}
\def\Author#1{\begin{center}{ \sc #1} \end{center}}
\def\Address#1{\begin{center}{ \it #1} \end{center}}

\newenvironment{Abstract}{\begin{center}{\bf Abstract}\end{center} \bigskip \begin{quotation}  }{\end{quotation}}
\newenvironment{Presented}{\begin{quotation} \begin{center} 
             PRESENTED AT\end{center}\bigskip 
      \begin{center}\begin{large}}{\end{large}\end{center} \end{quotation}}

\def\beq{\begin{equation}}
\def\eeq#1{\label{#1}\end{equation}}
\def\eeqn{\end{equation}}

\def\beqa{\begin{eqnarray}}
\def\eeqa#1{\label{#1}\end{eqnarray}}
\def\eeqan{\end{eqnarray}}

\let\bar=\overbar

\def\Dslash{\not{\hbox{\kern-4pt $D$}}}
\def\dslash{\not{\hbox{\kern-2pt $\del$}}}

\def\msb{{\bar{\ssstyle M \kern -1pt S}}}

\begin{document}
\begin{titlepage}

\vfill


\Title{$D^0-\overline{D}{}^0$ mixing and $CP$ Violation in charm}
\vfill
\Author{A. Zupanc\\ On behalf of the Belle Collaboration}  
\Address{Karlsruhe Institute of Technology, Karlsruhe, 76131, Germany}
\vfill


\begin{Abstract}
We review recent experimental results on $D^0-\overline{D}{}^0$ mixing and $CP$ violation charm decays. 
These studies provide complementary constraints on many different extensions of the Standard Model. 
Observation of $CP$ violation in charm decays at the current level of experimental sensitivity would 
be clear signals of New Physics.
\end{Abstract}

\vfill

\begin{Presented}
The Ninth International Conference on\\
Flavor Physics and CP Violation\\
(FPCP 2011)\\
Maale Hachamisha, Israel,  May 23--27, 2011
\end{Presented}
\vfill

\end{titlepage}
\def\thefootnote{\fnsymbol{footnote}}
\setcounter{footnote}{0}
%


\section{Introduction}

A process in which a particle changes to its antiparticle 
(flavor quantum number changes by two units) is called particle-antiparticle oscillation or mixing. 
Particle-antiparticle mixing has been observed in all four flavored neutral particle-antiparticle systems, 
i.e., in neutral kaon, both neutral $B$ meson systems and neutral $D$ meson system. 
The particle-antiparticle mixing phenomenon causes an initial (at time $t=0$), pure $D^0$ meson state 
(below expressions are written for neutral $D$ meson system, however they are the same in other three 
neutral meson systems that exhibit mixing) to evolve in time to a linear
combination of $D^0$ and $\overline{D}{}^0$ states:
\begin{equation}
 i\frac{d}{dt}
 \left (
   \begin{array}{c}
        D^0(t)\\
        \overline{D}{}^0(t)\\
   \end{array}
   \right )=\left[\mathbf{M}-\frac{i}{2}\mathbf{\Gamma}\right]\left (
   \begin{array}{c}
        D^0(t)\\
        \overline{D}{}^0(t)\\
   \end{array}
   \right ),
\end{equation}
where $\mathbf{M}$ and $\mathbf{\Gamma}$ are $2\times 2$ Hermitian matrices. 
Diagonal elements of the effective Hamiltonian $\mathbf{H}_{\rm eff}=\mathbf{M}-\frac{i}{2}\mathbf{\Gamma}$ describe 
flavor-conserving transitions $D^0\to D^0$ and $\overline{D}{}^0\to \overline{D}{}^0$, 
while off-diagonal elements describe the flavor-changing transitions 
$D^0\leftrightarrow\overline{D}{}^0$. The hermiticity of  $\mathbf{M}$ and $\mathbf{\Gamma}$ 
requires ${M}_{12}={M}^{\ast}_{21}$, ${M}_{ii}={M}^{\ast}_{ii}$, 
${\Gamma}_{12}={\Gamma}^{\ast}_{21}$ and ${\Gamma}_{ii}={\Gamma}^{\ast}_{ii}$, 
and the $CPT$ invariance requires ${M}_{11}={M}_{22}\equiv M$ and ${\Gamma}_{11}={\Gamma}_{22}\equiv \Gamma$.

The eigenstates of the effective Hamiltonian $\mathbf{H}_{\rm eff}$ are
\begin{equation}
 |D_{1,2}\rangle = p|D^0\rangle\pm q|\overline{D}{}^0\rangle,
\end{equation}
while the corresponding eigenvalues are
\begin{equation}
 \lambda_{1,2}=\left (M-\frac{i}{2}\Gamma\right)\pm\frac{q}{p}\left({M}_{12}-\frac{i}{2}{\Gamma}_{12}\right)\equiv m_{1,2}-\frac{i}{2}\Gamma_{1,2}.
\end{equation}
The coefficients $p$ and $q$ are complex coefficients, satisfying $|p|^2+|q|^2=1$, and
\begin{equation}
 \frac{q}{p}=\sqrt{\frac{M_{12}^{\ast}-\frac{i}{2}\Gamma_{12}^{\ast}}{M_{12}-\frac{i}{2}\Gamma_{12}}}=\left|\frac{q}{p}\right|e^{i\phi}.
\end{equation}
The real parts of the eigenvalues $\lambda_{1,2}$ represent masses, $m_{1,2}$, and their imaginary parts represent the widths $\Gamma_{1,2}$ of the two eigenstates $|D_{1,2}\rangle$, respectively. The time evolution of the eigenstates is given by 
\begin{equation}
 |D_{1,2}(t)\rangle = e_{1,2}(t)|D_{1,2}\rangle,~~~~~~e_{1,2}(t)=e^{-i(m_{1,2}-i\Gamma_{1,2}/2)t}.
\end{equation}

The time-dependent decay amplitudes of $D^0$ and $\dzerobar$ states decaying to final state $f$ are then given by
\begin{eqnarray}
 \langle f|{\cal H}|D^0(t)\rangle & = & \frac{1}{2}\left(\left[e_1(t)+e_2(t)\right]{\cal A}_f + \frac{q}{p}\left[e_1(t)-e_2(t)\right]\overline{\cal A}_f\right) \label{eq_th_tddr_d0f},\\
 \langle f|{\cal H}|\dzerobar(t)\rangle & = & \frac{1}{2}\left(\frac{p}{q}\left[e_1(t)-e_2(t)\right]{\cal A}_f + \left[e_1(t)+e_2(t)\right]\overline{\cal A}_f\right) \label{eq_th_tddr_d0barf},
\end{eqnarray}
where the instantaneous amplitudes ${\cal A}_f$ and $\overline{\cal A}_f$ are defined as 
${\cal A}_f\equiv\langle f|{\cal H}|D^0\rangle$, and $\overline{\cal A}_f\equiv\langle f|{\cal H}|\dzerobar\rangle$.
The time-dependent decay rates are obtained by squaring the above decay amplitudes:
\begin{eqnarray}
 \Gamma(D^0(t)\to f) & = & |{\cal A}_f|^2e^{-\Gamma t}\left( \frac{1+|\lambda_f|^2}{2}\cosh(y\Gamma t)-Re[\lambda_f]\sinh(y\Gamma t)\right.\nonumber\\
& & ~~~~~~~~~~~~\left. +\frac{1-|\lambda_f|^2}{2}\cos(x\Gamma t)+Im[\lambda_f]\sin(x\Gamma t)\right),\label{eq_th_dofdr}\\
\Gamma(\dzerobar(t)\to f) & = & |{\cal A}_f|^2\left|\frac{p}{q}\right|^2e^{-\Gamma t}\left( \frac{1+|\lambda_f|^2}{2}\cosh(y\Gamma t)-Re[\lambda_f]\sinh(y\Gamma t)\right.\nonumber\\
& & ~~~~~~~~~~~~~~~~~~~\left. -\frac{1-|\lambda_f|^2}{2}\cos(x\Gamma t)-Im[\lambda_f]\sin(x\Gamma t)\right),\label{eq_th_dobfbdr}
\end{eqnarray}
where $\lambda_f = \frac{q}{p}\frac{\overline{\cal A}_f}{{\cal A}_f}$, and $x$ and $y$ are dimensionless mixing parameters 
defined as
\begin{eqnarray}
 x & \equiv & \frac{m_1-m_2}{\Gamma},\label{eq_th_x}\\
 y & \equiv & \frac{\Gamma_1-\Gamma_2}{2\Gamma},\label{eq_th_y}
\end{eqnarray}
and $\Gamma\equiv(\Gamma_1+\Gamma_2)/2$ is the mean decay width.

In case of a non-zero mass difference ($x\neq 0$), the mixing is a consequence of a
 pure $D^0\leftrightarrow\overline{D}{}^0$ transition, and in the case of non-zero lifetime 
difference ($y\neq 0$), the mixing is a consequence of the shorter-living eigenstate dying out. 
Since the remaining longer-living eigenstate is a linear combination of weak eigenstates $D^0$ and $\overline{D}{}^0$, 
an initially pure $D^0$ sample produces some fraction of $\overline{D}{}^0$ over time. 

The time evolution of $\dzero,\dzerobar\to f$ decays is exponential with the lifetime $\tau_{\dzero}=1/\Gamma$, 
modulated by the mixing parameters $x$ and $y$ (see the expressions above). Time-dependent measurements of $\dzero$ 
and $\dzerobar$ decays thus enable us to measure the mixing parameters $x$ and $y$. 
Since the dependence on $x$ and $y$ is specific for each decay mode, 
different decay modes exhibit different sensitivities to the parameters $x$ and $y$.

\begin{figure}[t]
\begin{center}
 \includegraphics[width=0.45\textwidth]{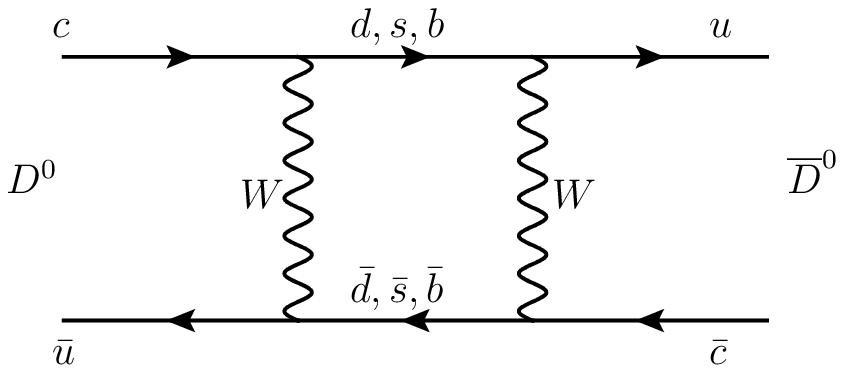}\hspace{0.09\textwidth}\raisebox{1cm}{\includegraphics[width=0.45\textwidth]{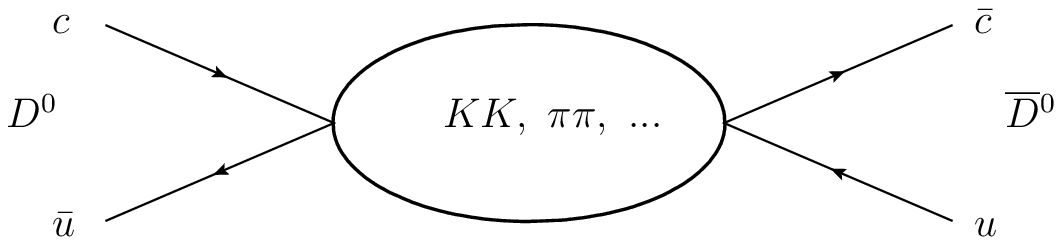}}
\end{center}
 \caption{Short distance (left) and long distance contributions \zdmix mixing in the Standard Model.}
 \label{fig_th_boxD0}
\end{figure}

Out of the four flavored neutral meson systems the neutral $D$ meson system is the only one in 
which down-type quarks are involved in the mixing loop (see Fig. \ref{fig_th_boxD0}). 
The neutral pion is its own antiparticle and top quark decays before it forms a hadron and therefore cannot oscillate. 
Studies of charm mixing offer therefore a
unique probe for New Physics (NP) via flavor changing neutral currents in the down-type quark sector.
In the Standard Model (SM) mixing in neutral $D$ meson system can proceed through a double weak boson exchange 
(short distance contributions) represented by box diagrams, or through intermediate states that are accessible to 
both $\dzero$ and $\dzerobar$ (long distance effects), as represented in Fig. \ref{fig_th_boxD0}. 
Potentially large long distance contributions are non-perturbative
and therefore difficult to estimate, so the predictions for the mixing parameters $x$ and $y$
within the SM span several orders of magnitude between $10^{-8}$ and $10^{-2}$~\cite{Nelson:1999fg,Falk:2001hx}. 
Due to large uncertainties of the SM mixing predictions it makes it difficult to identify NP 
contributions (clear hint would be, if $x$ is found to be much larger than $y$), however, 
measurements can still provide useful and competitive constraints on many NP models, as will be discussed later.

Study of $CP$ violation in decays of charmed hadrons also holds the potential for uncovering the NP. 
In the SM direct $CP$ violation can occur in singly Cabbibo suppressed (SCS; $c\to du\overline{d}$, $c\to su\overline{s}$) 
decays, but not in Cabbibo favored (CF; $c\to su\overline{d}$) or doubly Cabbibo suppressed (DCS; $c\to du\overline{s}$) 
decays. This is due to the fact that the final state particles in SCS decays contain at 
least one pair of quark and anti-quark of the same flavor, which makes a contribution 
from penguin-type or box amplitudes induced by virtual b-quarks possible 
in addition to the tree amplitudes. However, the contribution of these 
second order amplitudes are strongly suppressed by the small
combination of CKM matrix elements $V_{cb}V_{ub}^{\ast}$. The $CP$ violating asymmetry, defined as 
\begin{equation}
 A_{CP} = \frac{\Gamma(D\to f)-\Gamma(\overline{D}\to \overline{f})}{\Gamma(D\to f)+\Gamma(\overline{D}\to \overline{f})}
 \label{eq_ACP}
\end{equation}
is in the SM expected to be at most at the level of 0.1\% \cite{Buccella:1992sg}, which is well below the current experimental sensitivity. 
In some NP models the $CP$ asymmetry can be significantly enhanced and can be as large as 1\% \cite{Bigi:1994aw,Lipkin:1999qz,D'Ambrosio:2001wg,Grossman:2006jg}. 
It is thus widely believed that the observation of large $CP$ violation at the order of 1\% in charm decays would 
be an unambiguous sign for processes beyond the SM.
Direct $CP$ violation occurs when the absolute value of the decay amplitude for $D$ to decay to a final state $f$ (${\cal A}_f$)
is different from the one of corresponding $CP$-conjugated amplitude ($\overline{\cal A}_{\overline{f}}$). This can happen if the 
decay amplitude can be separated into at least two parts 
(in case of SCS decays the two corresponding SM amplitudes are the tree and the penguin-type) associated with 
different weak and strong phases, ${\cal A}_f=|{\cal A}_1|e^{i\delta_1}e^{i\phi_1}+|{\cal A}_2|e^{i\delta_2}e^{i\phi_2}$,
where $\phi_i$ represents weak phases that switch sign under $CP$-transformation, and $\delta_i$ represent strong phases 
which are $CP$-invariant. This ensures that $CP$-conjugated amplitude, $\overline{\cal A}_{\overline{f}}$, 
can differ from ${\cal A}_f$. Using the definition of $A_{CP}$ from above one obtains 
$A_{CP}\propto sin(\phi_1-\phi_2)sin(\delta_1-\delta_2)$. Experimentally, the direct $CP$-violation is probed by 
measuring the  difference between the partial decay widths ($\Gamma$) of $D\to f$ and $\overline{D}\to \overline{f}$.
In neutral $D$ meson system the time integrated $CP$ asymmetry, $A_{CP}$, can receive also contributions 
from $CP$ violation induced by mixing (if $|q/p|\neq 1$) or interference between mixing and decay (if $arg(q{\cal A}_f/p\overline{A}_f)\neq 0 (\pi)$ and $f=\overline{f}$).

\section{Recent experimental results on \zdmix mixing}

The most precise constraints on the mixing parameters $x$ and $y$ are obtained using the time dependence 
of $\dzero$ decays. In time-dependent measurements, the $\dzero$ decay time is calculated as
$t=m_{\dzero}(\vec{L}\cdot\vec{p}_{\dzero})/|\vec{p}_{\dzero}|^2$ , where $\vec{L}$ is the vector joining the $\dzero$'s 
production and decay vertices, and $\vec{p}_{\dzero}$ and $m_{\dzero}$ are its momentum and nominal mass. 
Detected tracks of $\dzero$ decay products are refitted to a common vertex to determine the $\dzero$ decay point, 
and the production point is taken to be at the primary vertex (in $p\overline{p}$  collisions) or is calculated 
from the intersection of $\dzero$ momentum vector with the beam spot profile (in $e^+e^-$ collisions). Often, the 
flavor of initially produced neutral $D$ mesons needs to be tagged in order to identify \zdmix transitions. 
The flavor is tagged by requiring that neutral $D$ mesons originate from $D^{\ast +}\to \dzero\pi^+$ decays. 
The charge of pion accompanying $\dzero$ tags the flavor of the neutral $D$ meson at production. 
The energy released in the $D^{\ast +}$ decay, $q=M_{D^{\ast +}}-M_{\dzero}-m_{\pi^+}$, has a narrow peak near the 
threshold for the signal events and thus helps also to reject the background candidates.

Mixing in the \zdmix system has been searched for more than two decades without success - until 2007. 
Three experiments - Belle, BaBar and CDF - so far have found evidence for this phenomenon. 
In the following, the measurements of the mixing parameters will be briefly summarized in different $\dzero$ meson decays.

\subsection{Decays to $CP$ eigenstates}

\begin{figure}[t]
\begin{center}
 \includegraphics[width=0.45\textwidth]{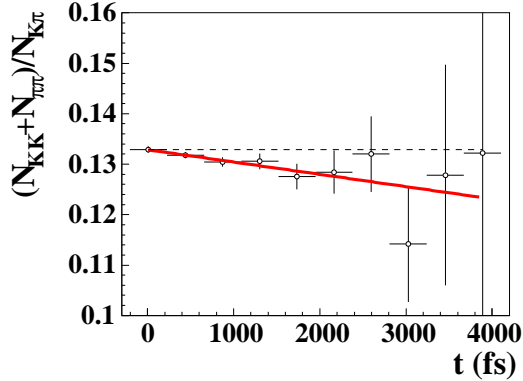}
\end{center}
 \caption{Proper decay time dependent ratio of $\dzero$ decays to $CP$-even eigenstates $K^-K^+$ and $\pi^-\pi^+$ to $CP$
 mixed final state $K^-\pi^+$ as measured by Belle \cite{Staric:2007dt}. The full (red) line shows the dependence of 
 this ratio for measured value of $y_{CP}$, while the dashed line shows expectation of no mixing ($y_{CP}=0$).}
 \label{fig_belle_ycp}
\end{figure}

Belle \cite{Staric:2007dt} found first evidence for \zdmix mixing using the ratios of lifetimes extracted from a 
sample of $\dzero$ mesons produced through the process $D^{\ast +}\to \dzero\pi^+$, which decay to $K^-\pi^+$, $K^-K^+$, or 
$\pi^-\pi^+$. The time-dependent decay rates of the CF mode, $K^-\pi^+$, and the SCS modes $h^-h^+$ ($h=K$ or $\pi$) 
are obtained from time-dependent decay rates given in previous section:
\begin{eqnarray}
 \Gamma(\dzero(t)\to K^-\pi^+,\dzerobar(t)\to K^+\pi^-) & \propto & e^{-t/\taudzero}\\
\Gamma(\dzero(t),\dzerobar(t)\to h^+h^-) & \propto & e^{-(1+y_{CP})t/\taudzero},
\end{eqnarray}
where it has been taken into account that $x,y\ll1$ and $|\overline{\cal A}_f/{\cal A}_f|=1$ 
($|\overline{\cal A}_f/{\cal A}_f|\ll1$) for $\dzero$ meson decays to $h^-h^+$ ($K^-\pi^+$). 
The lifetime difference between the $CP$ eigenstates $h^-h^+$ and CP-mixed state $K^-\pi^+$, $y_{CP}$, 
is defined as
\begin{equation}
 y_{CP}\equiv\frac{\tau_{K^{\mp}\pi^{\pm}}}{\tau_{h^+h^-}}-1=\frac{1}{2}\left(\left|\frac{q}{p}\right|+\left|\frac{p}{q}\right|\right)y\cos\phi-\frac{1}{2}\left(\left|\frac{q}{p}\right|-\left|\frac{p}{q}\right|\right)x\sin\phi.\label{eq_th_ycp}
\end{equation}
The lifetimes $\tau_{K\pi}$ and $\tau_{h^h}$ are the effective lifetimes extracted from the samples of $\dzero$ 
mesons decaying to $CP$ mixed final state $K^-\pi^+$, and $CP$ even final states $K^-K^+$ and $\pi^-\pi^+$. 
If $|q/p|=1$ and $\phi=0(\pi)$, the $CP$ symmetry in mixing and interference between mixing and decay is conserved, 
and the parameter $y_{CP}$ corresponds to the mixing parameter $y$. In these time-dependent measurements of 
neutral $D$ mesons decaying to $CP$ eigenstates the indirect $CP$ violation is also probed by comparing lifetimes 
of $\dzero$ and $\dzerobar$ mesons:
\begin{equation}
A_{\Gamma} = \frac{\tau(\dzero)-\tau(\dzerobar)}{\tau(\dzero)+\tau(\dzerobar)} .
\end{equation}
Using sample consisting of around 0.15 (1.2) million reconstructed tagged $\dzero$ decays to $h^-h^+$ ($K^-\pi^+$) Belle found 
$y_{CP} = (1.13\pm0.32\pm0.25)\%$ and $A_{\Gamma}=(0.01\pm0.30\pm0.15)\%$.
Figure \ref{fig_belle_ycp} shows proper decay time dependent ratio of $\dzero$ decays to $CP$-even eigenstates $K^-K^+$ and $\pi^-\pi^+$ to 
$CP$ mixed final state $K^-\pi^+$ as measured by Belle \cite{Staric:2007dt}. Results were later confirmed by BaBar's 
measurement using $\dzero$ tagged \cite{Aubert:2007en} and untagged \cite{Aubert:2009ck} samples. 

\subsection{Decays to hadronic wrong sign decays}

\begin{figure}[t]
\begin{center}
 \includegraphics[width=0.45\textwidth]{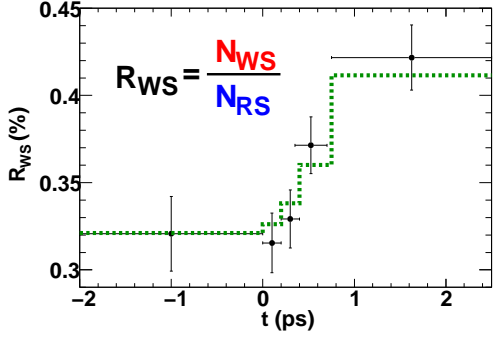}
\end{center}
 \caption{The ratio of wrong sign to right sign decays of $\dzero$ to $K^+\pi^-$ as measured by BaBar \cite{Aubert:2007wf}. 
The dashed line shows the expected wrong-sign rate as determined
from the mixing fit. In case of no mixing the ratio is expected to be constant. }
 \label{fig_babar_kpws}
\end{figure}

Babar found first evidence for \zdmix mixing in a time dependent study of DCS $\dzero\to K^+\pi^-$ decyas \cite{Aubert:2007wf}. 
These decays (also referred to as wrong sign decays) can precede both through mixing followed by a CF decay, 
$\dzero\to\dzerobar\to K^+\pi^-$, or directly through a DCS decay $\dzero\to K^+\pi^-$. 
To distinguish the two processes, an analysis of the decay time distribution is performed. The time-dependent decay 
rate for the two-body wrong sign decays $D^0\to K^+\pi^-$ is given by:
\begin{eqnarray}
 \Gamma(D^0(t)\to K^+\pi^-) & = & e^{-\Gamma t}|{\cal A}_{K^-\pi^+}|^2\left(R_D+\sqrt{R_D}\left|\frac{q}{p}\right|(y'\cos\phi-x'\sin\phi)\Gamma t\right.\nonumber\\
& & ~~~~~~~~~~~~~~~~~~~~~\left. +\frac{1}{4}\left|\frac{q}{p}\right|^2(x'^2+y'^2)(\Gamma t)^2\right),\label{eq_th_wsdo}\\
\end{eqnarray}
where $R_D$ is a ratio of DCS to CF decays, and the parameters $x'$ and $y'$ are rotated mixing parameters, 
rotated by an unknown strong phase difference between the DCS and CF amplitudes, $\delta_{K\pi}$: 
$x'=x\cos\delta_{K\pi}+y\sin\delta_{K\pi}$ and $y'=y\cos\delta_{K\pi}-x\sin\delta_{K\pi}$. 
The three terms in the time-dependent decay rates of wrong sign decays are due to the 
DCS amplitude, the interference between the DCS and CF amplitudes, and the CF amplitude, respectively. 
BaBar \cite{Aubert:2007wf} and CDF \cite{Aaltonen:2007uc} found evidence for oscillations in $\dzero\to K^+\pi^-$ 
(samples consisted of around 4 and 13 thousand wrong sign decays, respectively) 
with 3.9 and 3.8 standard deviations, respectively. 
The most precise measurement is from Belle \cite{Zhang:2006dp}, excluding the non-mixing point $x'^2=y'=0$, 
at 2.1 standard deviations. Figure \ref{fig_babar_kpws} shows proper decay time dependent wrong sign ratio 
($R_{WS} = N(\dzero\to K^+\pi^-)/N(\dzero\to K^-\pi^+)$) as measured by BaBar \cite{Aubert:2007wf}. Extraction of the mixing 
parameters $x$ and $y$ from the results of these measurements requires knowledge of the relative strong phase $\delta_{K\pi}$, 
which can be determined in time integrated measurements using the correlated \zdmix system produced at $\psi(3770)$ 
\cite{Asner:2008ft}.
Both, Belle and BaBar collaborations performed studies of $\dzero$ and $\dzerobar$ samples separately to probe the $CP$
violation in mixing. Results are found to be consistent with no $CP$ violation.

\subsection{Three-body decays}

\begin{figure}[t]
\begin{center}
 \includegraphics[width=0.45\textwidth]{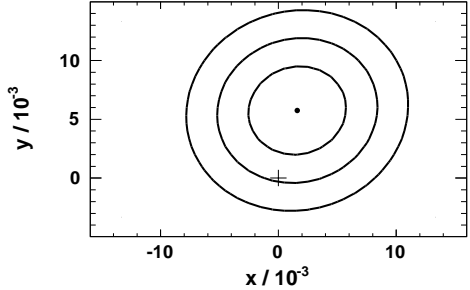}
\end{center}
 \caption{Central value (point) and C.L. contours 
(including statistical, systematic and amplitude model uncertainties) in the $x-y$ plane for 
C.L. = 68.3\%, 95.4\%, 99.7\% as obtained by BaBar in \cite{delAmoSanchez:2010xz}. The no-mixing point is shown as a plus sign ($+$) and is disfavored at 1.9 standard deviations. }
 \label{fig_babar_k0spipi}
\end{figure}

Several intermediate resonances can contribute to a hadronic three-body decay 
of a neutral $D$ meson. For example, $\dzero\to K^0_S\pi^+\pi^-$ decays can proceed 
via $D^0\to K^{\ast -}\pi^+$ (CF amplitude), $\dzero\to K^0_S\rho^0$ (SCS amplitude and $CP$ eigenstate), 
$\dzero\to K^{\ast +}\pi^-$ (DCS amplitude) and many others. 
In the isobar model, the instantaneous amplitudes for $\dzero$ and $\dzerobar$ decays to the three-body 
final state $f$ are parameterized as a sum of Breit-Wigner resonances and a constant non-resonant term 
(in case of no direct $CP$ violation, e.g. there is no difference between amplitudes and phases in $\dzero$ and $\dzerobar$
decays):
\begin{equation}
 {\cal A}_f(s_+,s_-) = \overline{\cal A}_f(s_+,s_-) = \sum_r a_re^{i\phi_r}{\cal A}_r(s_+,s_-) + a_{\rm NR}e^{i\phi_{\rm NR}},
\end{equation}
where $\sqrt{s_{\pm}}$ is the invariant mass of a pair of final state particles (e.g. $K^0_S\pi^{\pm}$), 
and sum runs over all resonances $r$. The time dependent decay rate for $\dzero$ decays is thus given 
by (similar expression is obtained for $\dzerobar$ decays):
\begin{eqnarray}
  \frac{d\Gamma(D^0\to f)}{ds_+ds_-dt} & \propto &
  |{\cal A}_1(s_+,s_-)|^2e^{-\frac{t}{\tau}(1+y)} + 
  |{\cal A}_2(s_+,s_-)|^2e^{-\frac{t}{\tau}(1-y)} \nonumber\\
  & & + 2Re[{\cal A}_1(s_+,s_-){\cal A}_2^{\ast}(s_+,s_-)]\cos{\left(x\frac{t}{\tau}\right)}
  e^{-\frac{t}{\tau}} \nonumber\\
  & & + 2Im[{\cal A}_1(s_+,s_-){\cal A}_2^{\ast}(s_+,s_-)]\sin{\left(x\frac{t}{\tau}\right)}
  e^{-\frac{t}{\tau}}, \label{eq_mD0}\\
\end{eqnarray}
where ${\cal A}_{1,2}(s_+,s_-) = \frac{1}{2}\left({\cal A}_f(s_+,s_-)\pm\frac{q}{p}\overline{\cal A}_f(s_+,s_-)\right)$. 
Different regions in the $s_+ - s_-$ plane (also called as the Dalitz plot) exhibit different forms of time 
dependence, as can be seen from the above decay rate; 
therefore, the time-dependent Dalitz plot analysis  of neutral $D$ meson decays to three-body final state 
enables us to measure the $x$ and $y$ parameters simultaneously. 
In case the analysis is performed separately for $\dzero$ and $\dzerobar$ samples $CP$ violation can be 
probed by measuring $q/p$ (amplitude and phase) directly. 
This method of measuring mixing parameter $x$ and $y$ was pioneered by Cleo in $\dzero\to K^0_S\pi^+\pi^-$ decays \cite{Asner:2005sz}, 
and applied also by Belle in $\dzero\to K^0_S\pi^+\pi^-$ decays \cite{Abe:2007rd}
and BaBar in $\dzero\to K^0_S\pi^+\pi^-$ and $\dzero\to K^0_SK^+K^-$ 
decays \cite{delAmoSanchez:2010xz}. Measurements are consistent with each other and 
together provide the most accurate determinations of $x$ and $y$ (average of all three measurements assuming no $CP$ violation, 
as obtained in Ref. \cite{Asner:2010qj}): $x=(0.42\pm0.21)\%$ and $y=(0.46\pm0.19)\%$.
Belle's (BaBar's) measurement disfavor the no-mixing hypothesis with 
a C.L. equivalent to 2.2 (1.9) standard deviations. Belle also performed search for $CP$
violation in mixing in their measurement and found $|q/p|=(0.86\pm0.30\pm0.09)$, $arg(q/p)=(-0.24\pm0.30\pm0.09)$,
to be consistent with 1 and 0, respectively. 
Figure \ref{fig_babar_k0spipi} shows the confidence level (C.L.) contours in the $x-y$ plane as obtained by BaBar collaboration.

Large fractions of $\dzero\to K^0_SK^+K^-$ decays
proceed via $\dzero\to K^0_S\phi$ (CP-odd) and $\dzero\to K^0_Sa_0(980)$ (CP-even) decays. 
Belle took advantage of this fact and performed Dalitz plot integrated measurement of $y_{CP}$ mixing
parameter using untagged sample of $\dzero\to K^0_SK^+K^-$ decays \cite{Zupanc:2009sy}.
They measure the effective lifetimes of $\dzero$ mesons, $\tau_{\rm ON,OFF}$, in two different regions of $K^+K^-$
invariant mass (at $\phi$ peak (ON) and $\phi$ sidebands (OFF)), which are given by 
$\tau_{\rm ON,OFF}=(1+(1-2f_{\rm ON,OFF})y_{CP})\taudzero$, where $f_{\rm ON,OFF}$ is $CP$-even fraction in ON or OFF region. 
The obtained value of $y_{CP}$ is consistent with $y$ obtained by Babar's Dalitz plot analysis \cite{delAmoSanchez:2010xz} 
using these decays only. The sensitivity to $y$ of both approaches is similar. 

\begin{figure}[t]
\begin{center}
 \includegraphics[width=0.24\textwidth]{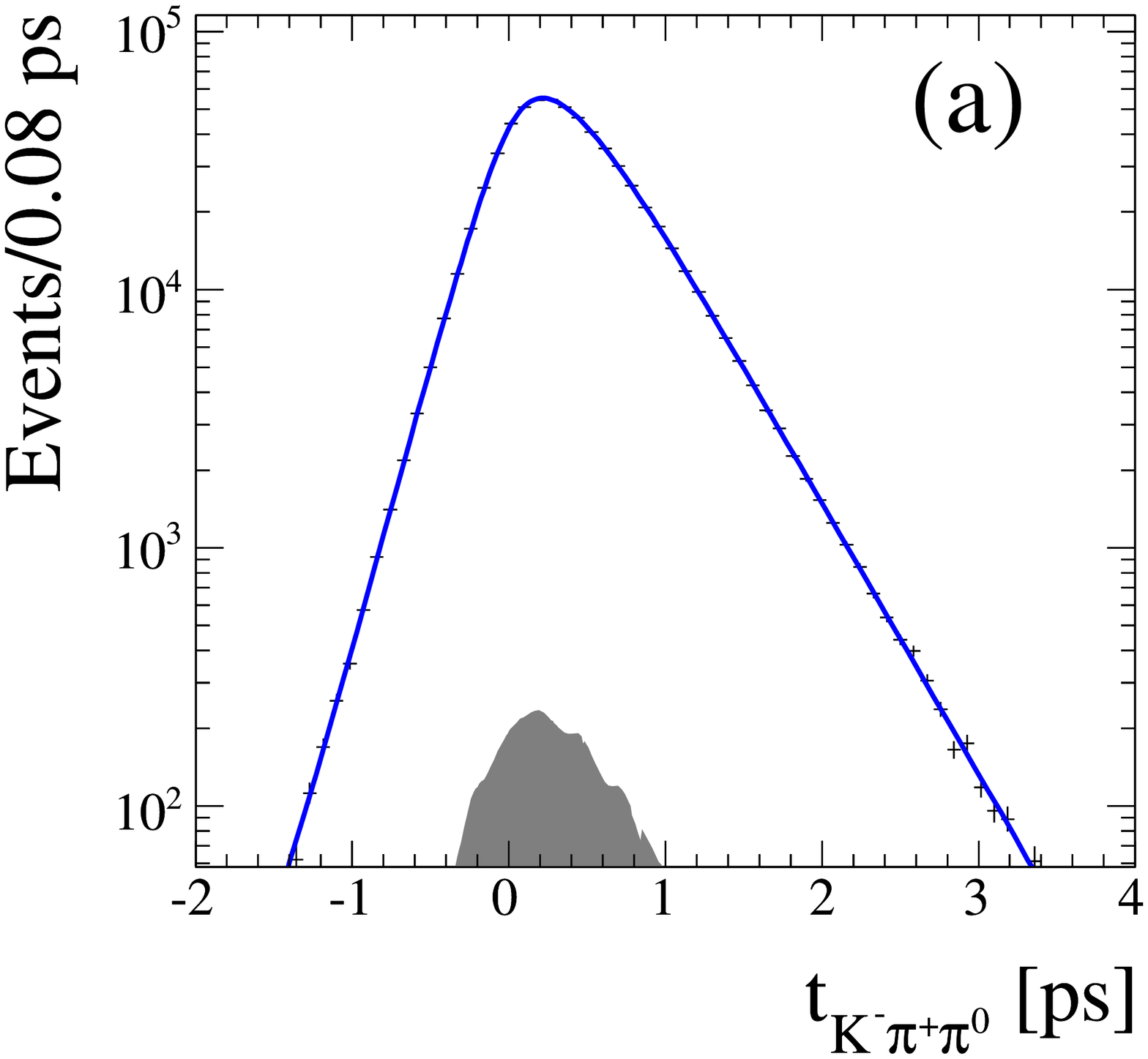}
 \includegraphics[width=0.24\textwidth]{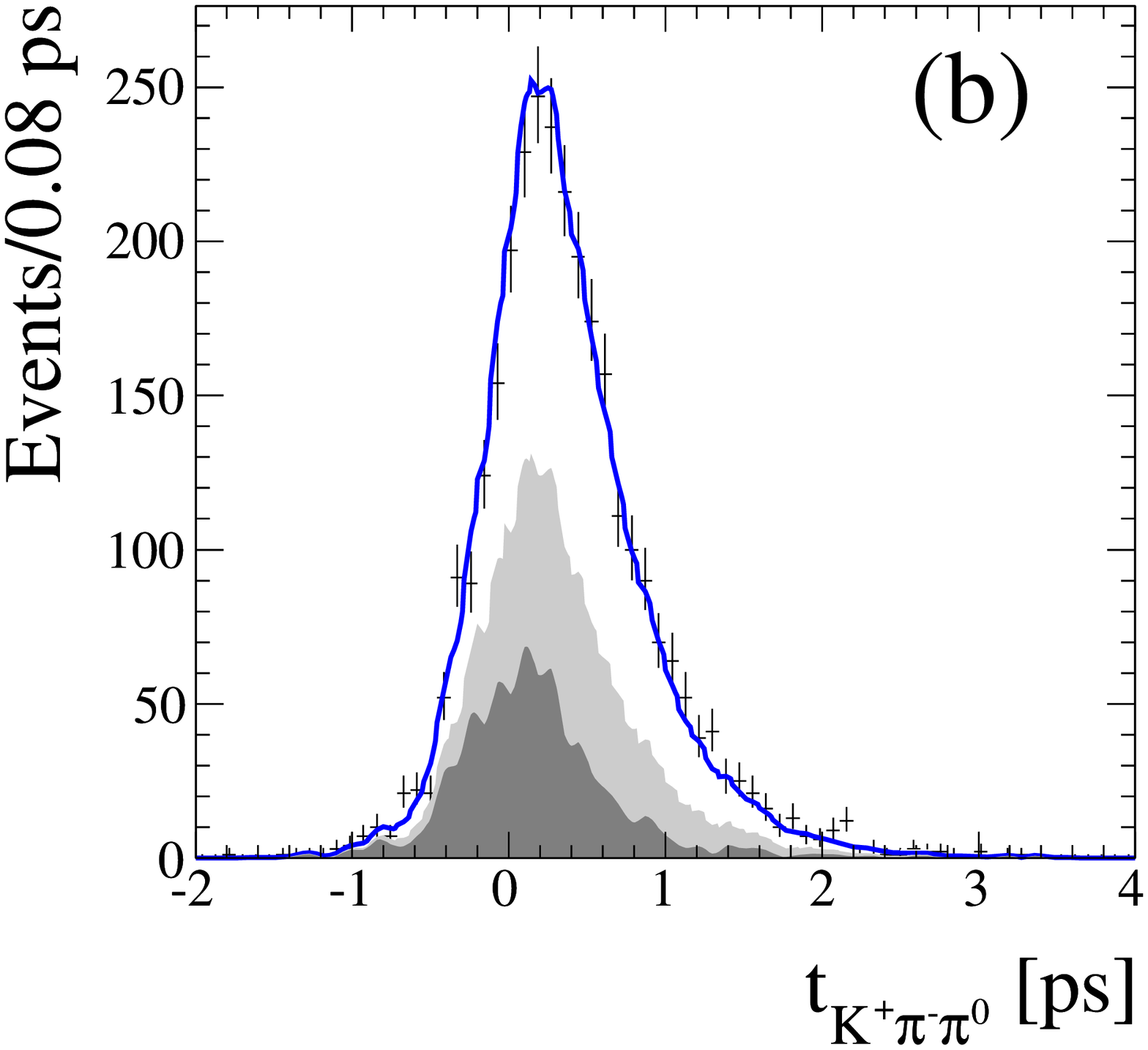}
\includegraphics[width=0.24\textwidth]{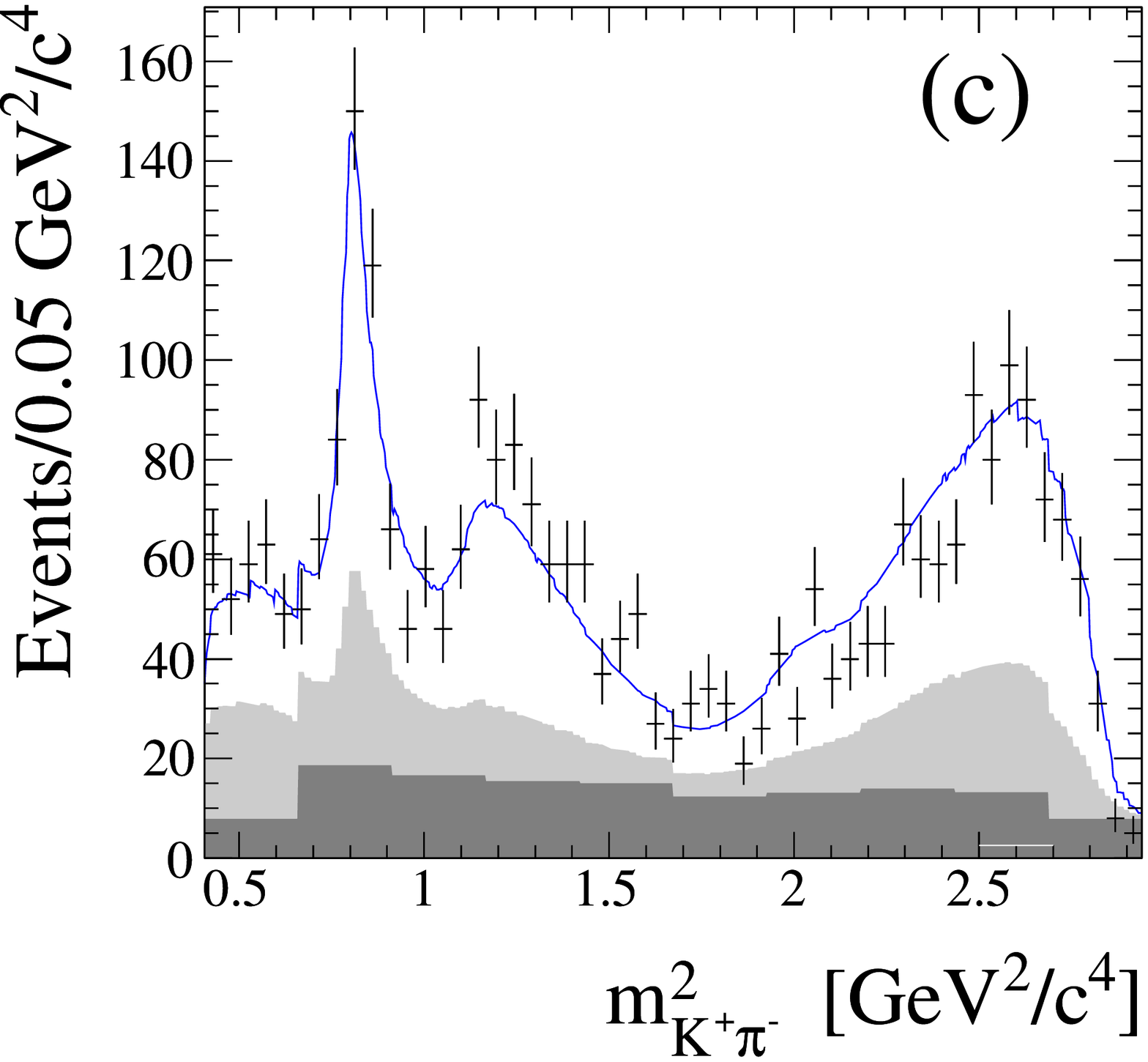}
\includegraphics[width=0.24\textwidth]{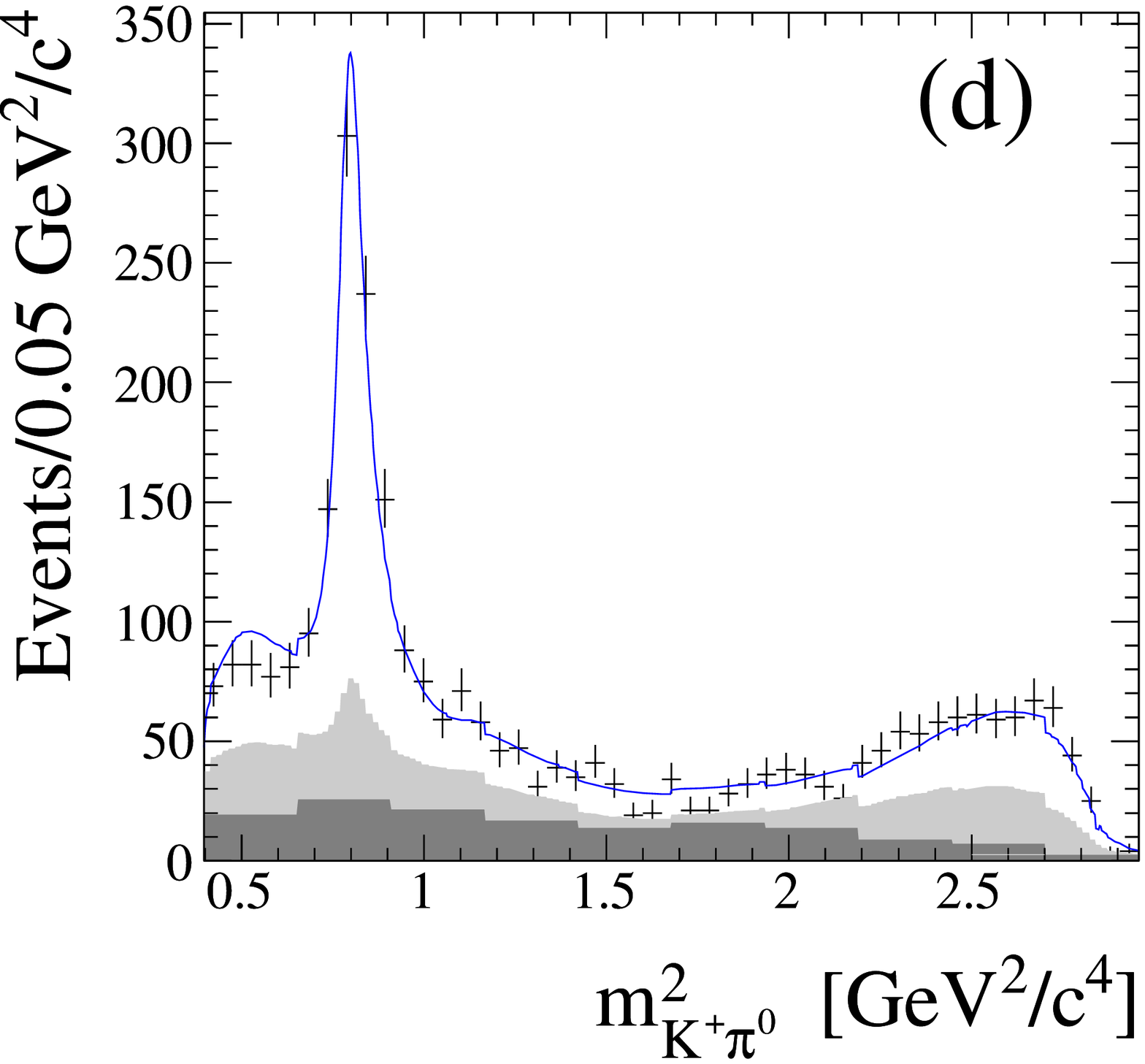}
\end{center}
 \caption{(a) Proper time distribution for right sign events with the fit result superimposed. 
The distribution of background events is shown by the shaded histogram. (b) Proper time distribution for wrong sign events. 
(c, d) $s_0$ and $s_+$ projections with superimposed fit results (line).}
 \label{fig_babar_kpipi0}
\end{figure}

BaBar found evidence of \zdmix mixing also with time-dependent analysis of wrong sign $\dzero\to K^+\pi^-\pi^0$ decays 
\cite{Aubert:2008zh}. Analysis is similar to analysis of wrong sign $\dzero\to K^+\pi^-$ decays, 
however since the strong phase difference, $\delta_{K\pi\pi^0}$, varies across the available 
three-body phase space the Dalitz plot analysis needed to be performed. 
The time dependent decay rate depends on the DCS amplitude ${\cal A}_{\overline{f}}$ and CF amplitude 
$\overline{\cal A}_{\overline{f}}$ and is given by:
\begin{eqnarray}
  \frac{d\Gamma(s_+,s_0,t)}{ds_+ds_0dt} & \propto & 
  \left \{ |{\cal A}_{\overline{f}}(s_+,s_0)|^2 
  + |{\cal A}_{\overline{f}}(s_+,s_0)\overline{\cal A}_{\overline{f}}(s_+,s_0)|
    (y cos\delta_{\overline{f}}-xcos\delta_{\overline{f}})t/\tau \right . \\\nonumber
  & & \left. +|\overline{\cal A}_{\overline{f}}(s_+,s_0)|^2\frac{x^2+y^2}{4}(t/\tau)^2\right \}e^{-t/\tau},
\end{eqnarray}
where $\delta_{\overline{f}}(s_+,s_0)=arg|{\cal A}^{\ast}_{\overline{f}}(s_+,s_0)\overline{\cal A}_{\overline{f}}(s_+,s_0)|$.
Figure \ref{fig_babar_kpipi0} shows projections of proper decay time and Dalitz plot variables $s_0$ and $s_+$ with 
superimposed fit results. The no-mixing point is excluded with a significance of 3.2 standard deviations.

\subsection{World average and constraints on New Physics models}

\begin{figure}[t]
\begin{center}
 \includegraphics[width=0.4\textwidth]{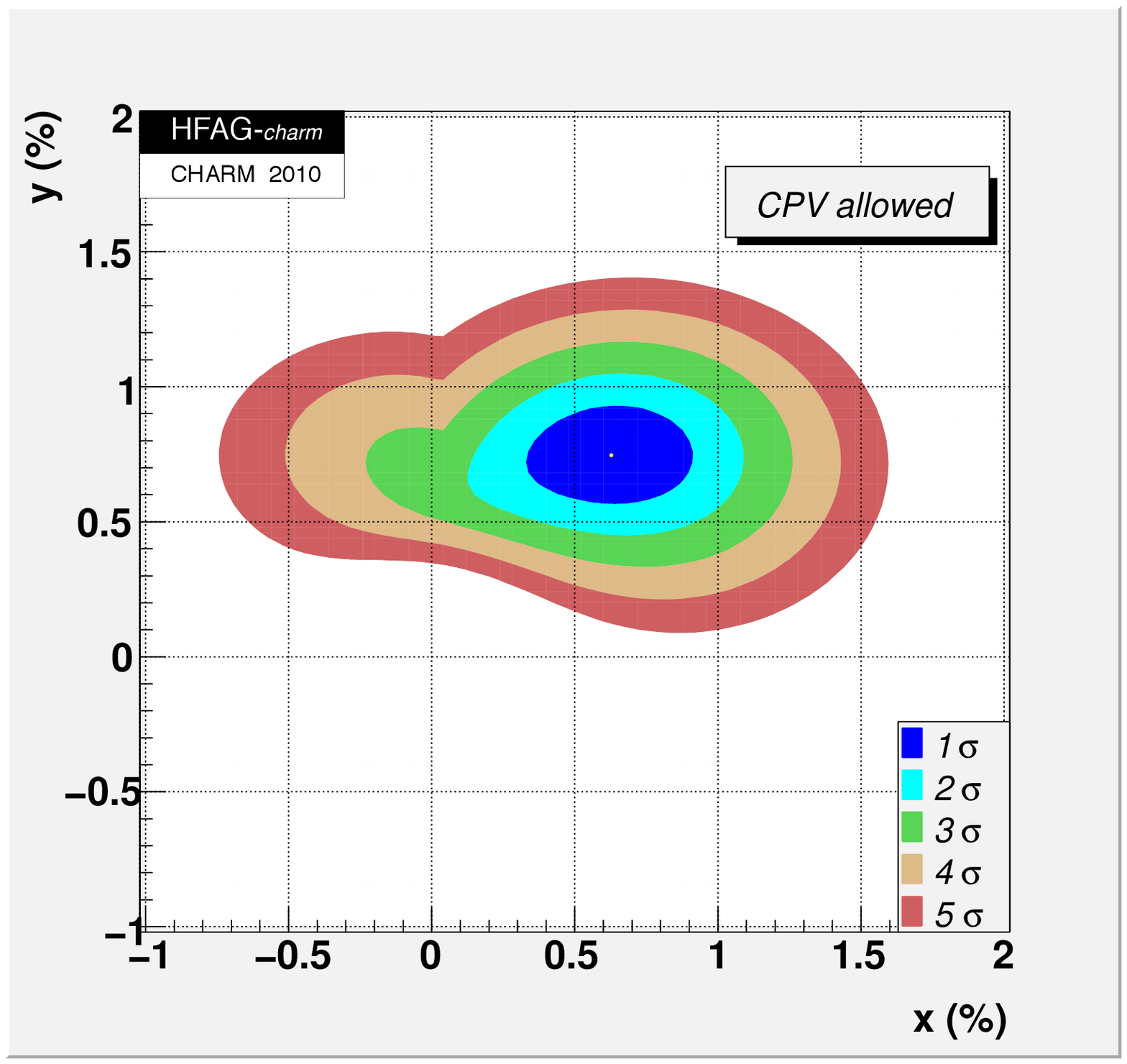}
 \includegraphics[width=0.4\textwidth]{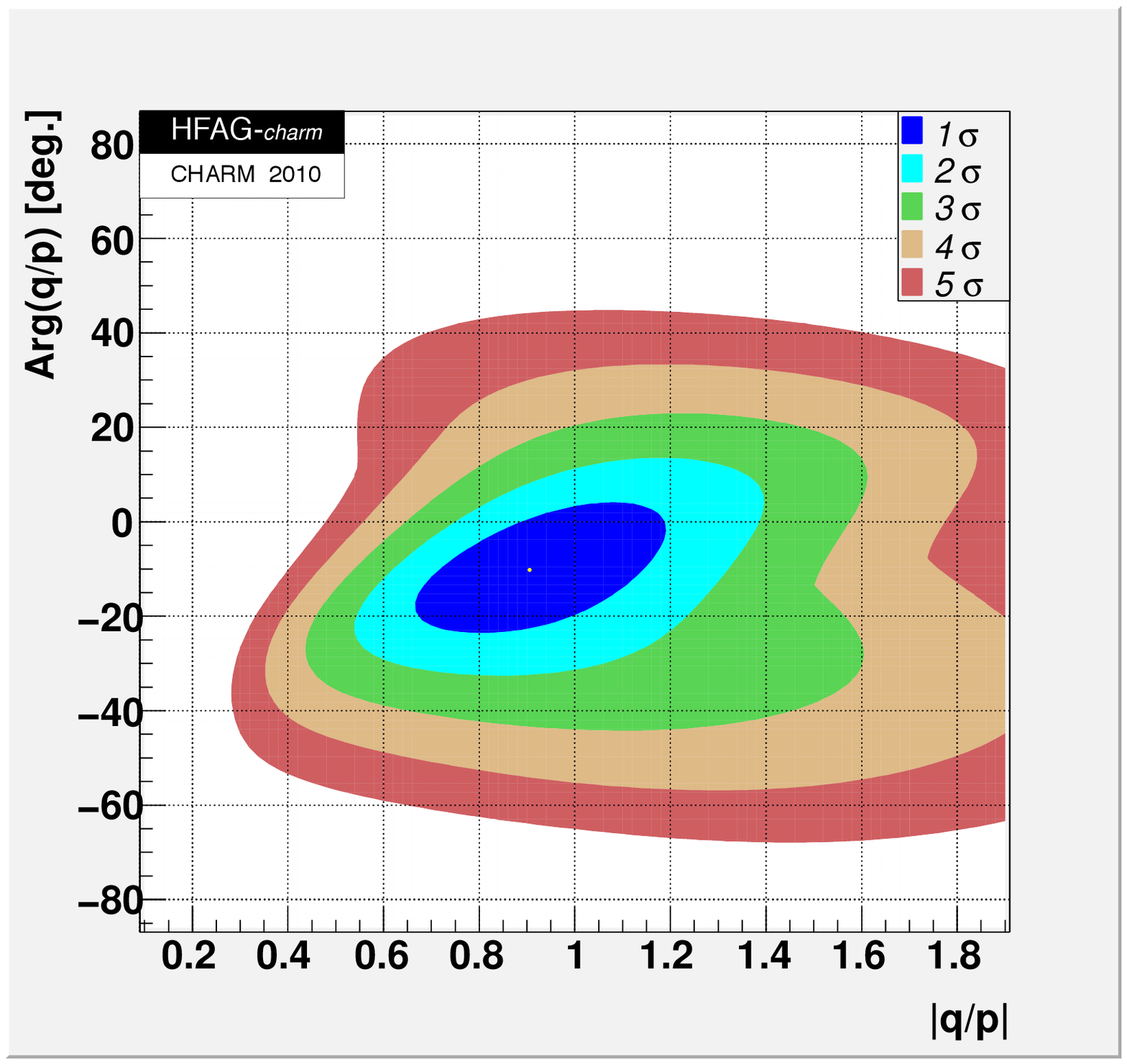}
\end{center}
 \caption{Two dimensional contours for mixing parameters $x$ and $y$ (left) and $CP$ violating parameters $|q/p|$ and $arg(q/p)$
 (right) as obtained by HFAG when averaging all relevant existing measurements.}
 \label{fig_mixwa}
\end{figure}

All existing measurements of \zdmix mixing (for full list of existing measurements check \cite{Asner:2010qj}) 
are combined by the Heavy Flavor Averaging Group by performing a global $\chi^2$ fit. The average central values 
are found to be $x=(0.63^{+0.19}_{-0.20})\%$ and $y=(0.75\pm0.12)\%$ \cite{Asner:2010qj}. Figure \ref{fig_mixwa} shows the 
resulting $1-5$ standard deviation contours in $x-y$ and $|q/p|-arg(q/p)$ planes. 
No \zdmix mixing hypothesis is excluded at 10.2 standard deviations, while there is no hint for $CP$ violation at 
current level of sensitivity. 

Golowich et al. \cite{Golowich:2007ka} studied implications of existing \zdmix measurements on many NP
models. They find in many scenarios strong constraints that surpass those from other search techniques 
and provide an important test of flavor changing neutral currents in the up-quark sector. 
One simple extension to the SM that they studied is the addition of a fourth family of fermions. 
The obtained constraint on the CKM mixing parameters $V_{cb'}V_{ub'}^{\ast}$ ($b'$ is the down-quark 
of the fourth generation) is an order of magnitude stronger than those obtained from unitarity 
considerations of the CKM matrix.

\section{Recent experimental results on direct $CP$ violation in charm}

In this section we will focus only on time-integrated measurements searching for direct $CP$ violation in decays of 
charmed hadrons since previous section already covered results from time-dependent studies. Searches of $CP$ violation 
were performed in past 15 years in over 30 decay modes of $\dzero$, $D^+$, and $D_S^+$ mesons by 
Belle, BaBar, Cleo, CDF, FOCUS, E796 and E687 experiments (the full list of all direct $CP$ violation measurements 
is available in Ref. \cite{Asner:2010qj}). No evidence for $CP$ violation were found so far, however 
the measurements have only started reaching interesting level of sensitivity below 1\% in some decay modes. 
In order to increase the sensitivity to or below 0.1\% level not only larger samples but also very good control 
over the systematic uncertainties will be needed. These uncertainties are dominated by the uncertainties in asymmetries 
in the detection and reconstruction of particles of opposite charge. In addition, forward-backward production asymmetries, 
$A_{FB}$, resulting from $Z^0$ and virtual photon interference and higher order loops in the production of $c$-anti-$c$ 
quark pairs in $e^+e^-$ collisions, results in asymmetries in the distribution of $D$ decay products in regions of 
varying efficiency in the detector. 
Estimation of these factors used to rely upon Monte Carlo simulated studies, with questionable assumptions about 
charge dependent interaction effects, resulting in systematic uncertainties in $A_{CP}$'s in the $1-5$\% range. 
In the past years, new insights in using real data rather than simulations have led to reduction of these uncertainties 
to the $0.2-0.5$\% range. These corrections and systematic uncertainties are decay mode dependent, 
however it is important to note that are determined by the statistics of the samples used and will thus decrease with 
increasing sample sizes.

\subsection{Direct $CP$ violation in neutral $D$ meson decays to $CP$ eigenstates}

Time-integrated $CP$ asymmetry, $A_{CP}$, defined in Eq. \ref{eq_ACP} receives in case of neutral $D$ meson decays 
 contributions of direct and indirect (mixing induced) $CP$ violation. The latter is independent of the decay mode 
(universal) and its time-dependence is for decays to $CP$ eigenstate given
by \cite{Du:2006jc}:
\begin{equation}
 A_{CP}(t)\approx \frac{1}{2} \eta_{CP} \left( y\left( |p/q| - |q/p| \right)cos\phi + x\left( |p/q| + |q/p| \right)sin\phi \right) \frac{t}{\tau}.
\end{equation}
Integrating over time yields
\begin{equation}
 A_{CP} = a^{\rm dir}_{CP} + \int A_{CP}(t)D(t)dt \approx a^{\rm dir}_{CP} + \frac{\langle t \rangle}{\tau}a^{\rm ind}_{CP}. 
\label{eq_acptime}
\end{equation}
The factor $\frac{\langle t \rangle}{\tau}$ in front of the second term, indirect $CP$ violation term, is 1, if the 
reconstruction efficiency does not depend on proper decay time, $D(t)=1$ (e.g. as it is the case at Belle and Babar). At CDF and LHCb the 
displaced-track trigger requirements reject candidates with short decay times which results in $\frac{\langle t \rangle}{\tau}>1$.

\begin{figure}[t]
\begin{center}
 \includegraphics[width=0.4\textwidth]{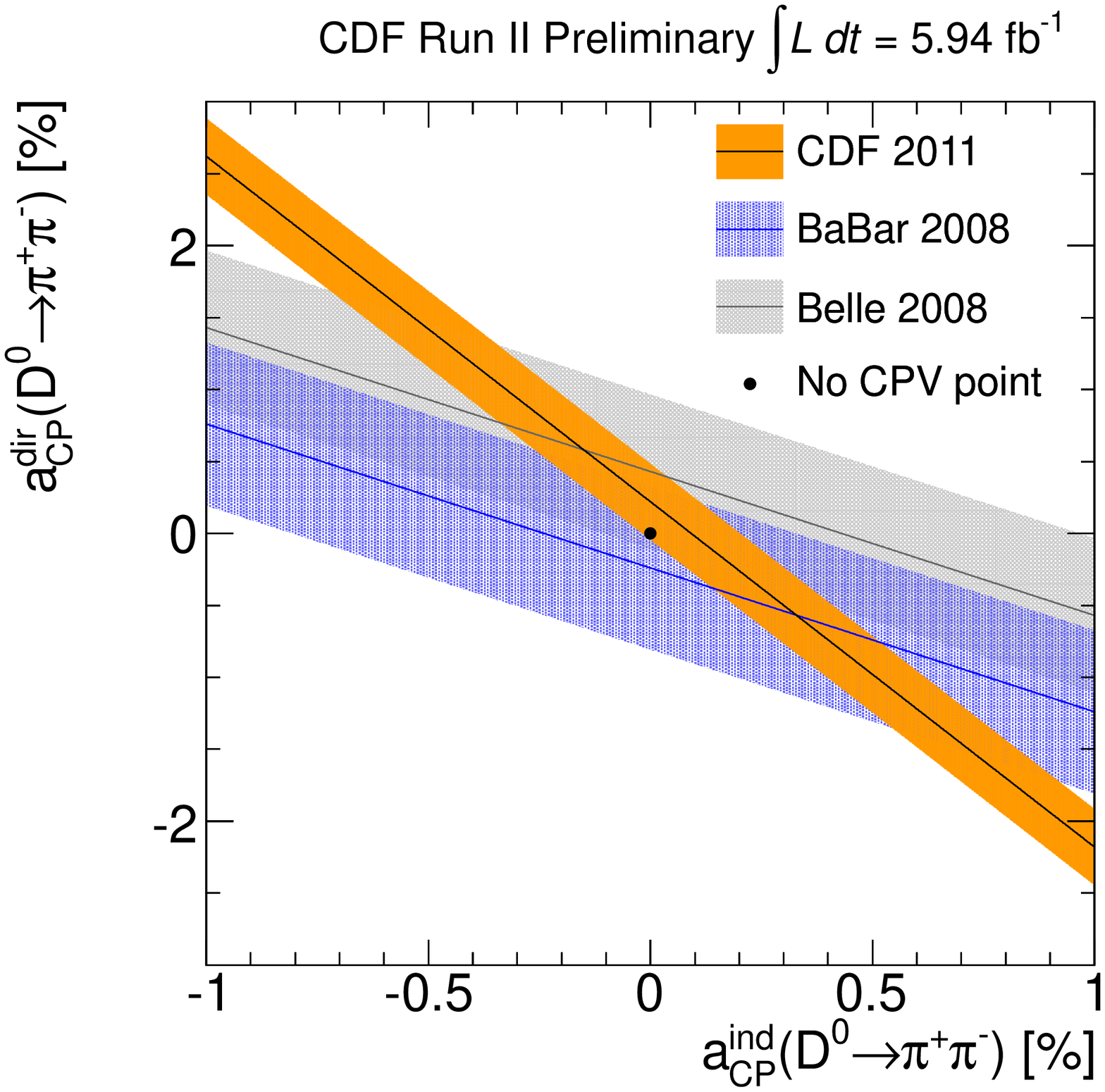}
 \includegraphics[width=0.4\textwidth]{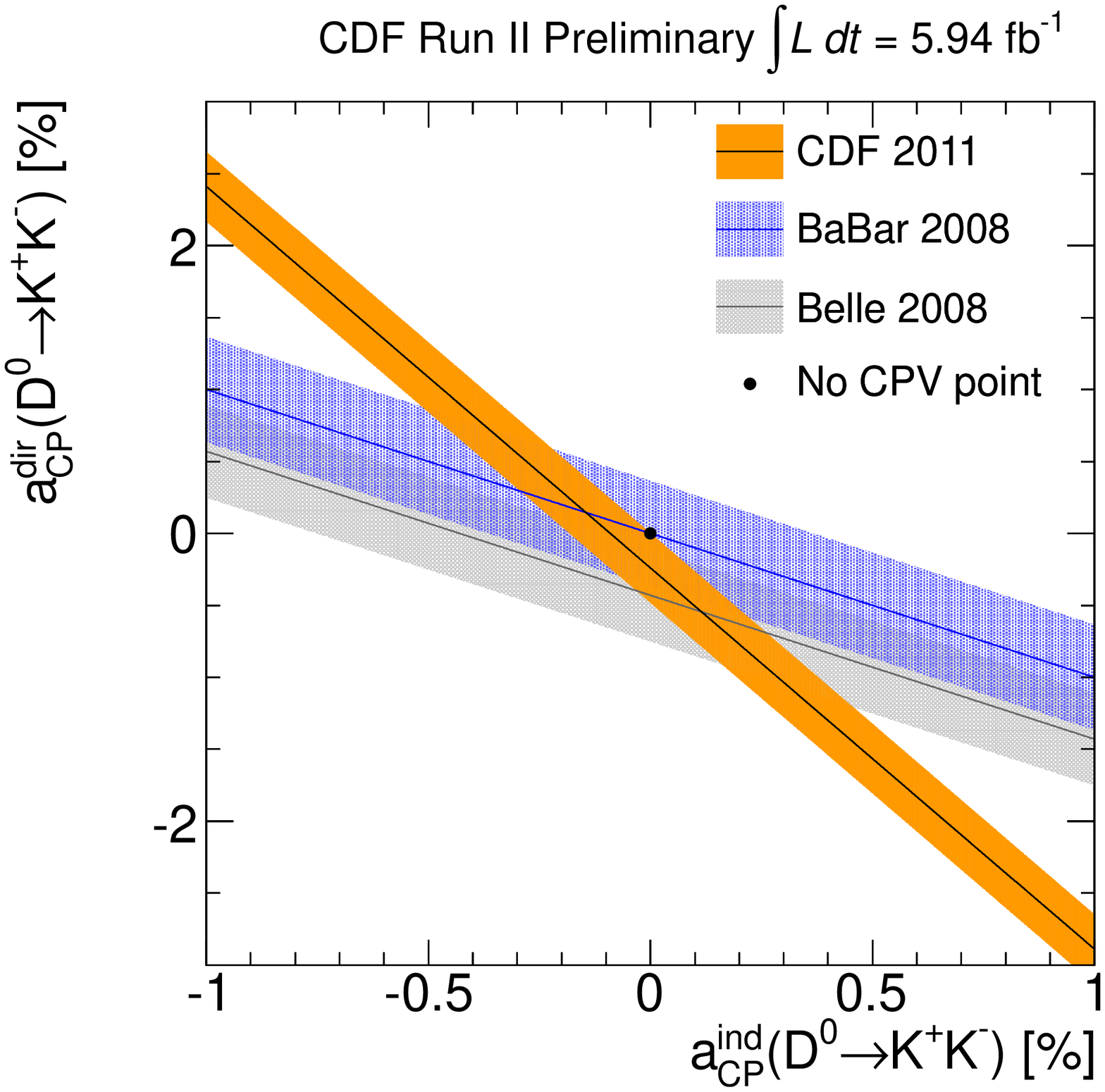}
\end{center}
 \caption{Comparison of $A_{CP}(\pi\pi)$ (left) and $A_{CP}(KK)$ (right) in the parameter space 
($a_{CP}^{\rm ind}$,$a_{CP}^{\rm dir}$) measured by CDF, Belle and Babar collaborations. }
 \label{fig_acpcdf}
\end{figure}

Recently, CDF presented measurements of $A_{CP}$ in neutral $D$ meson decays to $CP$ eigenstates $K^-K^+$ and $\pi^-\pi^+$ 
\cite{CDFacp}. The $D^0$ mesons are required to originate from $D^{\ast+}\to\dzero\pi^+$ decays. The raw reconstructed
asymmetry, defined as $A^{\rm reco}(hh)=(N(D^0\to hh)-N(\dzerobar\to hh))/(N(D^0\to hh)+N(\dzerobar\to hh))$, is a 
sum possible $CP$ asymmetry in $D^0$ decays and detector induced slow pion reconstruction efficiency asymmetry:
\begin{eqnarray}
 D^{\ast+}\to \dzero\pi_s^+\to [h^+h^-]\pi_s^+ & \Rightarrow & A^{\rm reco}(hh)^{\ast} = A_{CP}(hh) + A_{\varepsilon}(\pi_s).
\end{eqnarray}
In order to correct for the latter asymmetry and to extract the $A_{CP}(hh)$ asymmetry CDF reconstructed also tagged and untagged
samples of $\dzero\to K^-\pi^+$ decays. The raw reconstructed asymmetries in these two samples are given by:
\begin{eqnarray}
 D^{\ast+}\to \dzero\pi_s^+\to [K^-\pi^+]\pi_s^+ & \Rightarrow & A^{\rm reco}(K\pi)^{\ast} = A_{CP}(K\pi) + A_{\varepsilon}(\pi_s) + A_{\varepsilon}(K\pi),\\
 \dzero\to K^-\pi^+ & \Rightarrow & A^{\rm reco}(K\pi)\phantom{^{\ast}} = A_{CP}(K\pi) \phantom{+ A_{\varepsilon}(\pi_s)}~ + A_{\varepsilon}(K\pi).\\
\end{eqnarray}
The asymmetry $A_{CP}(hh)$ is then obtained by the following combination of raw reconstructed asymmetries:
\begin{equation}
 A_{CP}(hh)=A^{\rm reco}(hh)^{\ast}-A^{\rm reco}(K\pi)^{\ast}+A^{\rm reco}(K\pi),
\end{equation}
assuming that the production asymmetry to be 0 (which is the case since at Tevatron the initial state, $p\overline{p}$, is charge
symmetric), the efficiency to reconstruct $D^{\ast}$ can be factorized into $\pi_s$ and $\dzero$ reconstruction efficiencies and
that kinematic equations are equal across all four samples. The CDF collaboration found,
\begin{eqnarray}
 A_{CP}(\pi\pi)^{\ast}&=&(+0.22\pm0.24\pm0.11)\%,\\
 A_{CP}(KK)^{\ast}&=&(-0.24\pm0.22\pm0.10)\%,
\end{eqnarray}
which are consistent with $CP$ conservation. These are the most precise measurements up to date of time-integrated asymmetries of $D^0$ decays
 to $K^-K^+$ and $\pi^-\pi^+$ final states. Figure \ref{fig_acpcdf} shows the comparison of this measurement with the ones performed by
Belle and Babar collaborations \cite{Staric:2008rx,Aubert:2007if} in the parameter space ($a_{CP}^{\rm ind}$,$a_{CP}^{\rm dir}$) 
(according to Eq. \ref{eq_acptime}). 

\begin{figure}[t]
\begin{center}
 \includegraphics[width=0.4\textwidth]{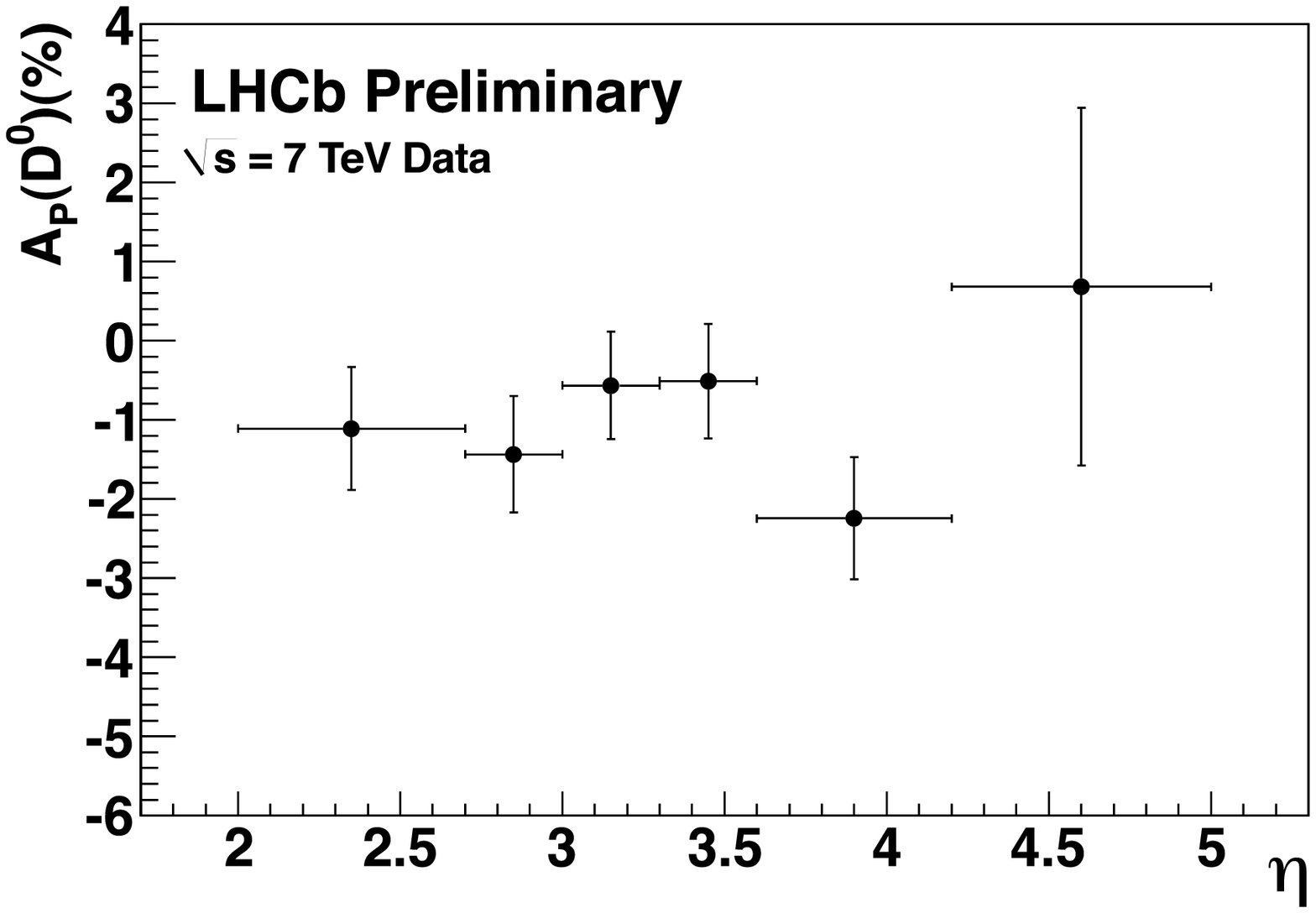}
 \includegraphics[width=0.4\textwidth]{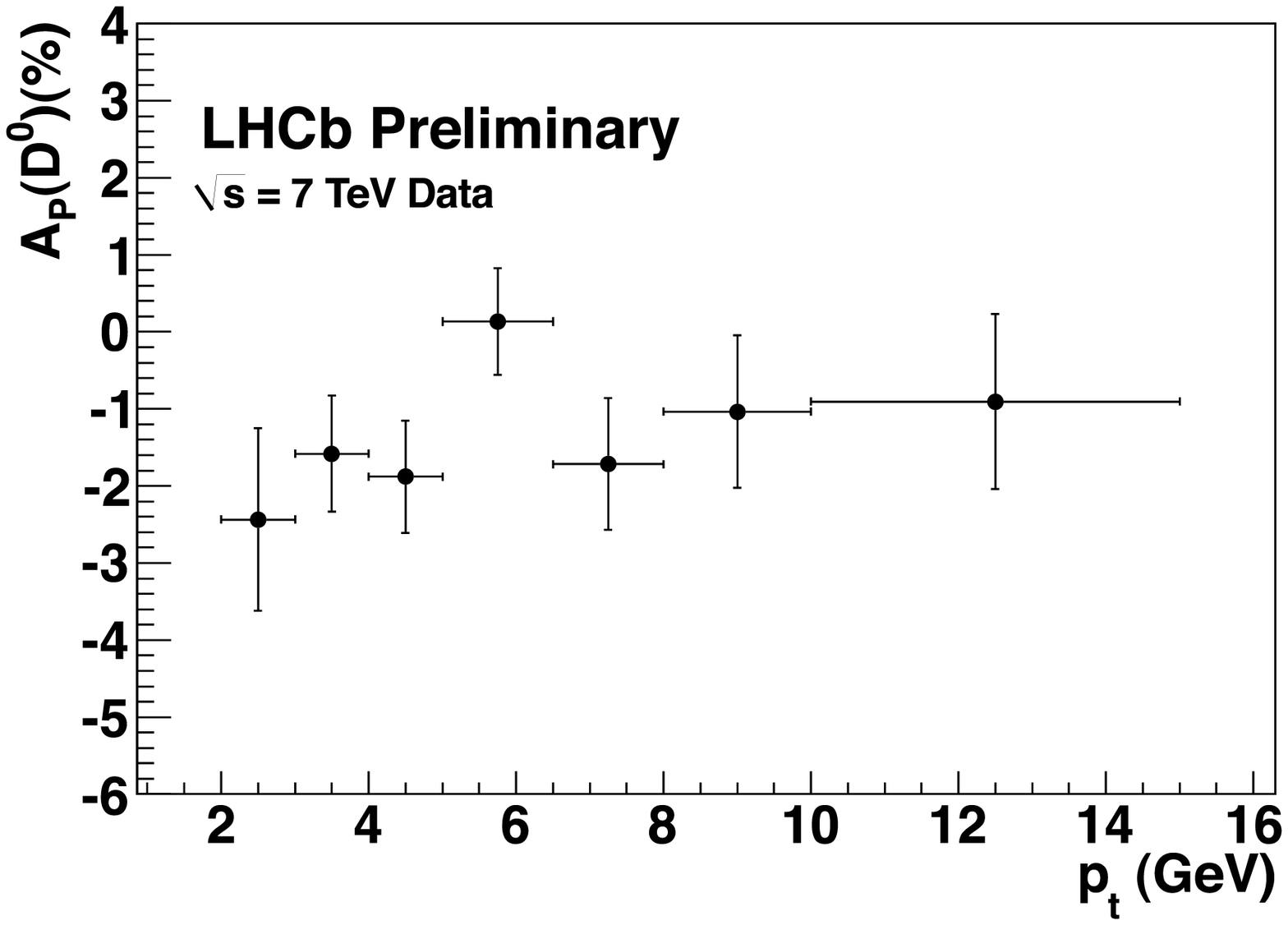}
\end{center}
 \caption{Measured production asymmetry of $D^0$ mesons in bins of $\eta$ (left) and $p_T$ (right). }
 \label{fig_aprodlhcb}
\end{figure}

Recently the LHCb collaboration presented a measurement of the difference of the time-integrated $CP$ asymmetries
in $D^0\to K^-K^+$ and $D^0\to \pi^-\pi^+$ decays using data collected in 2010 \cite{LHCbacp} in which production
asymmetry and detector related asymmetries cancel:
\begin{eqnarray}
 A^{\rm reco}(KK)^{\ast}&=&A_{CP}(KK)^{\ast}+A_{\varepsilon}(\pi_{\rm s})+A_P(D^{\ast}),\\
 A^{\rm reco}(\pi\pi)^{\ast}&=&A_{CP}(\pi\pi)^{\ast}+A_{\varepsilon}(\pi_{\rm s})+A_P(D^{\ast}),\\
\Delta A_{CP} = A_{CP}(KK)^{\ast}-A_{CP}(\pi\pi)^{\ast}& =& A^{\rm reco}(KK)^{\ast} - A^{\rm reco}(\pi\pi)^{\ast}.
\end{eqnarray}
They measured $\Delta A_{CP} = (-0.28\pm0.70\pm0.25)\%$ to be consistent with 0. Using also tagged and untagged samples of
$D^0\to K^-\pi^+$ decays the production
asymmetry of $D^{0}$ mesons, $A_P(D^0)$, can be expressed and measured if world averages for $A_{CP}(hh)$ are
taken as external inputs. Figure \ref{fig_aprodlhcb} shows the measured $D^{\ast}$ production asymmetry as a function of transverse 
momentum, $p_T$, and pseudo-rapidity, $\eta$. At current level of statistics the production asymmetry does not significantly
depend on $p_T$ and $\eta$, but when averaged out over LHCb's acceptance region is inconsistent with 0.

\subsection{Direct $CP$ violation in decays of charged $D$ mesons}

\begin{figure}[t]
\begin{center}
 \includegraphics[width=0.6\textwidth]{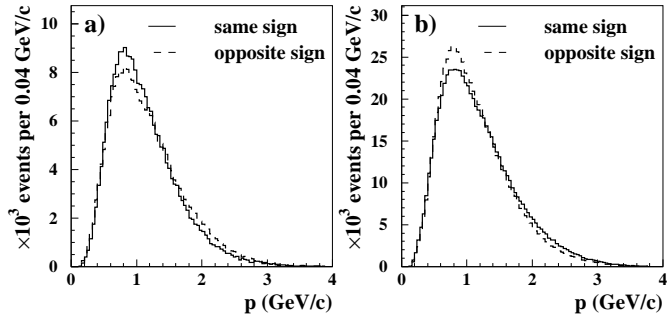}
\end{center}
 \caption{Background subtracted momentum distributions of same and opposite charge relative to that of 
  the $D$ (a) or $D_s$ (b) decays. }
 \label{fig_kkmom}
\end{figure}

\begin{figure}[t]
\begin{center}
 \includegraphics[width=0.4\textwidth]{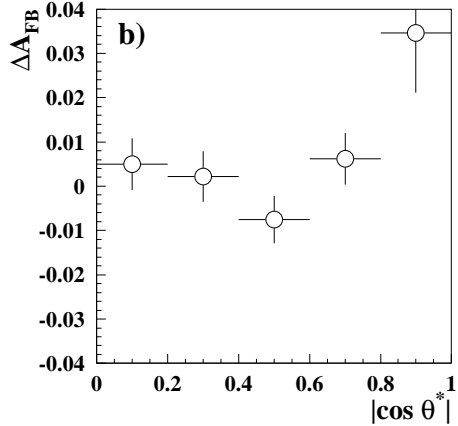}
\end{center}
 \caption{Forward-backward asymmetry difference in bins of $cos\theta^{\ast}$ between $D^+$ and $D^+_s$ mesons. No significant difference is observed. }
 \label{fig_dafb}
\end{figure}

Recently, Belle collaboration presented results of the measurement of $A_{CP}$ asymmetry difference between 
SCS $D^+\to\phi\pi^+$ and CF $D_s^+\to\phi\pi^+$ decays in the region of $|M(K^+K^-)-m_{\phi}|<16$~MeV/$c^2$ \cite{belledphipi}. 
In case of charged $D_{(s)}$ decays to $\phi\pi^+$ the
reconstructed asymmetry can be divided into four contributions
\begin{equation}
 A^{\rm reco} = A_{CP} + A_{FB}(cos\theta^{\ast})+A_{\varepsilon}(K^+K^-)+A_{\varepsilon}(p_{\pi},cos\theta_{\pi}),
\label{eq_phipi}
\end{equation}
where $A_{FB}(cos\theta^{\ast})$ is the forward-backward production asymmetry of the $D_{(s)}$ mesons, 
$A_{\varepsilon}(K^+K^-)$ is the asymmetry in the reconstruction of $K^+K^-$ pair, and 
$A_{\varepsilon}(p_{\pi},cos\theta_{\pi})$ is the asymmetry in the pion detection. Naively, one would
expect the $A_{\varepsilon}(K^+K^-)$ to be zero, since the $K^+K^-$ pair is charge symmetric. This would be the case,
if the $K^+$ and $K^-$ momentum spectra would be identical. As can be seen in Fig. \ref{fig_kkmom} the momentum spectra
are different which leads to non-zero $A_{\varepsilon}(K^+K^-)$. This asymmetry can be very precisely determined using 
$A_{\varepsilon}(K)$ (measured on data with $D_s^+\to\phi\pi^+$ and $D^0\to K^-\pi^+$ decays) and normalized
kaon phase space distributions $P(x)$ ($x=(p,cos\theta)$): 
\begin{eqnarray}
A_{\varepsilon}(K^+K^-)&=&\int(P_{K^+}(x)-P_{K^-}(x))A_{\varepsilon}(K)dx,\\
\Delta A_{\varepsilon}(K^+K^-) & = & A_{\varepsilon}^D(K^+K^-) - A^{D_s}_{\varepsilon}(K^+K^-) =(+0.111\pm0.025)\%.
\end{eqnarray}
The difference in reconstructed asymmetries is according to Eq. \ref{eq_phipi} then given by
\begin{equation}
 \Delta A^{\rm reco} = A_{CP}(D^+\to\phi\pi^+)+\Delta A_{FB}(cos\theta^{\ast})+\Delta A_{\varepsilon}(K^+K^-),
\end{equation}
where the intrinsic $A_{CP}$ in CF $D^+_s\to\phi\pi^+$ is assumed to be negligible and provided the measurement is
performed in bins of the $(cos\theta^{\ast}, p_{\pi}, cos\theta_{\pi})$ phase space.
To obtain $A_{CP}(D^+\to\phi\pi^+)$ and $\Delta A_{FB}(cos\theta^{\ast})$ the 
weighted average of $\Delta A^{\rm reco}$ in each bin of $cos\theta^{\ast}$ is calculated and then 
$A_{CP}(D^+\to\phi\pi^+)$ and $\Delta A_{FB}(cos\theta^{\ast})$ are extracted by adding or subtracting bins at 
$\pm cos\theta^{\ast}$. The $CP$ asymmetry in $D^\to\phi\pi^+$ decays is found to be $(0.51\pm0.28\pm0.05)\%$, 
and is consistent with no direct CP violation in these decays. This is the most precises measurement of $A_{CP}$ in these
decays up to date. In addition no significant difference between forward-backward asymmetries in the production of the 
$D^+$ and $D_s^+$ mesons is found as can be seen in Fig. \ref{fig_dafb}.

\section{Conclusions}
Studies of charm mixing and $CP$ violation offer unique opportunities to search for
processes beyond the SM. Large samples of neutral $D$ meson collected by Belle, BaBar and CDF 
experiments enabled us to find evidence for $D$ meson mixing after 31 years of its discovery. 
Next round of experiments (LHCb at CERN, BelleII at KEK in Japan, and SuperB in Italy) will collect 
even orders of magnitude larger samples which might reveal also $CP$ violation in charm. 
The future was never more bright for charm physics.

\end{document}